\documentclass[twocolumn,prl,superscriptaddress]{revtex4-2} 
\usepackage[T1]{fontenc}
\usepackage{amsmath,amsfonts,bm}

\usepackage{bbm}

\usepackage{amssymb}
\usepackage{graphicx}  %epsfig,
\usepackage{color}
\usepackage{extarrows}
\usepackage{enumitem}
\usepackage{empheq}
\usepackage{tensor}
\usepackage{orcidlink}
\usepackage[ruled, vlined]{algorithm2e}

 \usepackage[scr=boondoxupr]{mathalfa} 
\newcommand{\la}{\langle}
\newcommand{\ra}{\rangle}

\newcommand{\be}{\begin{equation}}
\newcommand{\ee}{\end{equation}}
\newcommand{\bsube}{\begin{subequations}}
\newcommand{\esube}{\end{subequations}}
\newcommand{\Eq}[1]{Eq.\,(\ref{#1})}

\newcommand{\Fig}[1]{Fig.\,\ref{#1}}

\newcommand{\RN}[1]{%
  \textup{\uppercase\expandafter{\romannumeral#1}}%
}

\usepackage{siunitx}

\allowdisplaybreaks[1]

\definecolor{darkblue}{RGB}{0, 56, 102}

\usepackage{hyperref}  
\hypersetup{hidelinks,
	colorlinks=true,
	allcolors=black,
	pdfstartview=Fit,
	breaklinks=true
}

\begin{document}

\title{Unit-Circle Moment Closure}

%%%

% \affiliation{
%   Hefei National Laboratory,  University of Science and Technology of China, Hefei, Anhui 230088, China
% }

\author{Yu Su}
\email{suyupilemao@mail.ustc.edu.cn}
\affiliation{
  Hefei National Research Center for Physical Sciences at the Microscale, University of Science and Technology of China, Hefei, Anhui 230026, China
}

\author{Yao Wang}
% \email{wy2010@ustc.edu.cn}
\affiliation{
  Hefei National Research Center for Physical Sciences at the Microscale, University of Science and Technology of China, Hefei, Anhui 230026, China
}

% \author{Wenjie Dou}
% \affiliation{
%   Department of Physics and Department of Chemistry, Westlake University, Hangzhou, Zhejiang 310024, China
% }

\date{\today}

\begin{abstract}

Moment closure is a central problem in reduced descriptions of stochastic, kinetic, and quantum dynamics, where equations for low-order observables are coupled to an unresolved hierarchy of higher-order moments. Existing closures usually impose a prescribed form on the distribution or directly truncate the hierarchy, which can become inaccurate or unstable for strongly non-Gaussian states. Here we introduce unit-circle moment closure, which recasts the problem as analytic continuation. Raw moments are mapped to bounded unit-circle moments, whose unresolved tail is reconstructed by a Takagi--Prony procedure from the effective pole structure of a mapped generating function. The resulting continuation yields stable higher-order moments without assuming a fixed distributional ansatz. Illustrative static and dynamical examples demonstrate accurate reconstruction of non-Gaussian distributions and stable evolution of moment hierarchies. Our approach provides a general perspective for moment closure based on analytic structure rather than direct truncation.

\end{abstract}

\maketitle

\paragraph*{Introduction.}
Moment methods provide a reduced description for a continuous and nonnegative density $\rho(x)$. Instead of dealing with the full distribution, one considers a hierarchy of its moments. However, this approach faces a basic difficulty. A finite set of moments does not, in general, determine a unique probability distribution \cite{Akh20,Mea842404}. For example, the truncated sequence \(\{\mu_0,\cdots,\mu_N\}\), where
\begin{align}\label{def_moments}
	\mu_n=\int x^n \rho(x)\,\mathrm{d}x
\end{align}
can be produced by many other different measures. The corresponding densities share the same low-order moments, but may have different higher-order moments, tails, or peak structures. This connects moment closure to the classical truncated moment problem \cite{Akh20,Mea842404}: given a finite moment sequence, one asks whether it can arise from a nonnegative density and, if so, how the sequence may be extended to higher orders.

The moment-closure problem adds dynamics to this ambiguity \cite{Kue15,Kue16,Gra49331,Lev961021,Ran06178101,Ott10120601,Ale13174101,Fro25063022,Alv26025207}. In a moment hierarchy, the equation for lower-order moments usually depends on the higher-order ones. Numerically evolving the equations of motion therefore needs a rule for estimating \(\{\mu_{N+1},\mu_{N+2},\cdots\}\) from \(\{\mu_0,\cdots,\mu_N\}\). From this perspective, moment closure is a structured extension of a truncated moment sequence. Such an extension should preserve physical realizability, remain numerically stable, and be simple enough to compute. Moment closure is therefore more than a direct truncation of the hierarchy. It selects one continuation from many distributions that agree with the known low-order moments.

Moment method is applied in many areas of physics and chemistry. In kinetic theory, velocity moments of the Boltzmann equation give equations for density, flow velocity, pressure, and heat flux \cite{Hua87}. Each equation depends on moments of higher order, so a closure is needed to obtain a finite set of hydrodynamic equations. A similar problem appears in radiative transfer and plasma physics \cite{Alv26025207}, where angular or velocity moments determine energy densities and fluxes, while higher-order fluxes remain unknown. Moment hierarchies also appear in stochastic reaction networks. For nonlinear reactions, the equations for mean populations depend on correlations, and the equations for correlations depend on higher-order fluctuations \cite{Sch15185101,Lev961021,Ran06178101,Ott10120601,Ale13174101,Wuy22054312}. In nonequilibrium statistical mechanics, the short-time expansion coefficients of a correlation function are moments of its spectral density \cite{Liu25148001,Flo20557277}. A finite set of such moments contains useful information about the spectrum, but does not uniquely determine its full shape or long-time dynamics. Similar hierarchical structures arise in quantum many-body theory, where reduced density matrices of successive particle order are coupled through the Bogoliubov--Born--Green--Kirkwood--Yvon (BBGKY) hierarchy \cite{Hua87}.

% Different closure schemes use different selection rules. Gaussian and cumulant closures assume that sufficiently high connected moments vanish. Maximum-entropy closures choose the least informative distribution that matches the retained moments. Quadrature-based methods approximate the distribution by a finite set of effective nodes and weights. These methods work well when the true distribution is close to the assumed form. Their accuracy can decrease for sharp peaks, well-separated modes, or strong non-Gaussian tails. The problem is even more severe for raw moments on an unbounded domain, because high-order moments can grow rapidly and become poorly conditioned.

Several methods have been developed for the moment closure problem. Cumulant closure assumes that sufficiently high connected moments vanish \cite{Lev961021,Gri12154105}. Polynomial expansions approximate the density in a chosen basis \cite{Gra49331,Bli98193}. They are simple, but may produce oscillations and may violate positivity. Pad\'e-like methods approximate the measure by a finite sum of point masses \cite{Bre0287,Gor68655,Sta71663}. They are closely related to Stieltjes continued fractions and Lanczos procedures \cite{Liu26174118}. Maximum-entropy methods choose the least-biased positive density consistent with the known moments, usually in the form $\rho_N(x)\propto\exp(-\sum_{n=1}^{N}\lambda_n x^n)$ \cite{Mea842404,Ran06178101}. These methods are useful, but each of them has a built-in bias. Cumulant method is useful when the distribution is close to Gaussian. Polynomial expansions favor polynomial densities. Pad\'e-like methods favor discrete measures. Maximum entropy favors exponential-polynomial densities. 

\begin{figure*}[ht]
	\centering
	\includegraphics[width=0.85\textwidth]{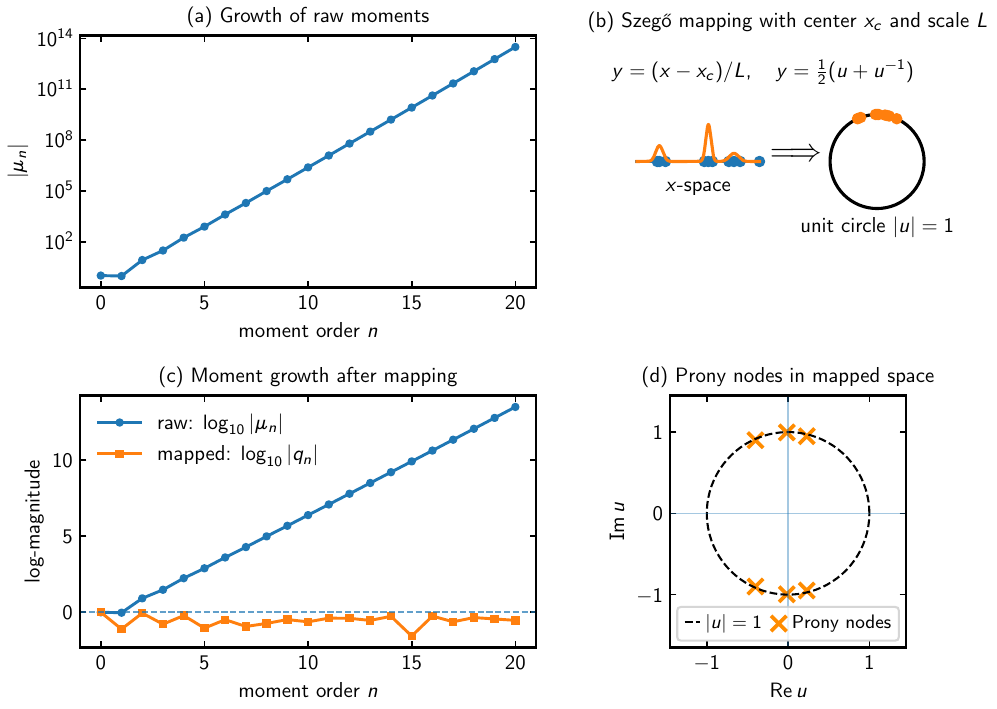}
	\caption{Schematic illustration of the unit-circle moment closure. Here, we demonstrate with the density $\rho(x) = \sum_{i}\alpha_i\varphi_i(x)$, where $\varphi_i(x)$ are Gaussian functions with different centers and widths, and $\{\alpha_i\}$ are positive weights. (a) The raw moments $\mu_n=\int x^n\rho(x)\,\mathrm{d}x$ grow rapidly with the moment order $n$, making direct continuation unstable. (b) The physical variable is shifted and scaled as $y=(x-x_c)/L$, and the finite interval is mapped to the unit circle through the $y=(u+u^{-1})/2$. This gives the Chebyshev representation $T_n(y)=(u^n+u^{-n})/2$. (c) After this transformation, the mapped moments $q_n=\int T_n((x-x_c)/L)\rho(x)\,\mathrm{d}x$ remain bounded and are much better conditioned than the raw moments.
(d) The tail of the mapped sequence is reconstructed by a Prony-type procedure. The extracted six Prony roots lie on or near the unit circle and encode the dominant modes used to continue the unresolved moment sequence.
}\label{fig1}
\end{figure*}

% Also, raw moments may grow very fast with the order $n$, which makes direct continuation in the monomial basis unstable.

In this work, we propose a different continuation scheme, called unit-circle moment closure. The main idea is to avoid continuing the raw moments [i.e., \Eq{def_moments}] directly. We first map the raw moments to a bounded sequence of unit-circle moments by using the Szeg{\H{o}} mapping \cite{Sim05,Sim13,Sil974822,Zha24035154,Zha25214111}. Then we reconstruct the unresolved tail of this new sequence as a Prony-type spectral problem \cite{Bey0517}. Finally, we map the continued unit-circle moments back to raw moments. 

This construction separates two difficulties in moment problem. The Szeg{\H{o}} mapping controls the growth of the moment sequence. It replaces the monomial basis by bounded modes on the unit circle. The Prony step then extracts the dominant spectral modes that can be resolved from the finite data. Thus the method does not assume a density of a fixed form, such as a polynomial, a finite sum of delta functions, or an exponential of a polynomial. Instead, it constructs a finite spectral continuation of the moment sequence on a compact complex domain. 
% In \Fig{fig1}, we illustrate the method with a concrete example, where the density is a mixture of three Gaussian functions. The raw moments grow rapidly with $n$, while the unit-circle moments remain bounded and well conditioned. The Prony procedure then extracts the dominant modes that are used to continue the tail of the moment sequence.
%  Dynamical moment hierarchies are a natural application of this idea. 

\paragraph*{Unit-circle moments.}
We now define the transformed moments. Choose a center $x_c$ and a scale $L$, and introduce the dimensionless variable $y=(x-x_c)/L$. For a density supported on, or mainly concentrated in, the region $|y|\leq 1$, we use the Szeg{\H{o}} parametrization \cite{Sim05,Sim13,Zha24035154,Zha25214111}
\begin{equation}\label{joukowski_map}
y=\frac{1}{2}\left(u+u^{-1}\right),\quad |u|=1 .
\end{equation}
This maps the physical interval to the unit circle.
The corresponding polynomial basis is the Chebyshev basis, since $T_n(y)=(u^n+u^{-n})/2$ on $|u|=1$ is the $n$-th Chebyshev polynomial. We therefore define the unit-circle moments as
\begin{equation}\label{unit_circle_moment}
q_n \equiv \int T_n\left(\frac{x-x_c}{L}\right)\rho(x)\,\mathrm{d}x .
\end{equation}
These moments are directly related to the raw moments defined in \Eq{def_moments}. Since $T_n(y)$ is a polynomial of degree $n$, $q_n$ only depends on $\mu_0,\mu_1,\cdots,\mu_n$.
Thus we can write
\begin{equation}\label{q_mu_transform}
q_n=\sum_{k=0}^{n}C_{nk}\mu_k .
\end{equation}
The coefficients $\{C_{nk}\}$ are fixed once $x_c$ and $L$ are chosen. They are obtained by expanding $T_n((x-x_c)/L)$ in powers of $x$. Therefore, the known raw moments $\mu_0,\cdots,\mu_N$ determine the known unit-circle moments $q_0,\cdots,q_N$. The transformation is triangular and has a nonzero diagonal part. Thus it can also be inverted order by order. After the unit-circle moments are continued, we can map them back to the original moments.

The main advantage of this representation is numerical stability. The monomial $x^n$ strongly amplifies the tail of a distribution. As a result, the raw moments may grow very fast with $n$. In contrast, $|T_n(y)|\leq 1$ for $|y|\leq 1$. Therefore, for a normalized density mainly supported in the mapped interval, the sequence ${q_n}$ behaves like Fourier coefficients on the unit circle. The moment continuation problem is then moved from an ill-conditioned monomial basis to a bounded spectral basis. In \Fig{fig1}, we illustrate our strategy with a concrete example, where the density is a mixture of Gaussian functions. In the panel (c) of \Fig{fig1}, the raw moments grow rapidly with $n$, while the unit-circle moments remain bounded and well conditioned.

This unit-circle representation also gives a natural complex-variable picture. By applying the Szeg{\H{o}} mapping, the interval in the real-axis is represented by the contour $|u|=1$. The unit-circle moments may be viewed as contour moments of a mapped analytic function. This is the reason why the continuation of ${q_n}$ can be written as a Prony problem.

\paragraph*{Residue representation and Takagi--Prony continuation.}
% Let $G(u)$ be the mapped analytic object whose contour moments give the unit-circle moments.
% Up to convention-dependent factors from the Jacobian and the symmetric combination of $u^n$ and $u^{-n}$, we write
With $x=x_c+L\cos\theta$, the unit-circle moments become $q_n=\int_0^\pi g(\theta)\cos(n\theta)\,\mathrm{d}\theta$ with $g(\theta)=L\rho(x_c+L\cos\theta)\sin\theta$. Thus $q_n$ is a cosine Fourier coefficient of the mapped density. By extending $g(\theta)$ evenly to $[0,2\pi]$ and setting $u=e^{i\theta}$, the same coefficient can be written as a unit-circle contour moment \cite{Zha24035154,Zha25214111},
\begin{equation}\label{contour_qn}
q_n=\frac{1}{2\pi i}\oint_{|u|=1}u^nG(u)\,\mathrm{d}u .
\end{equation}
where $G(u)$ denotes the analytic continuation of the symmetrized mapped density, including the Jacobian of the change of variables. \Eq{contour_qn} is the complex-analytic version of \Eq{unit_circle_moment}. It shows that $q_n$ probes the singularity structure of $G(u)$ inside the unit circle.

Suppose that the dominant structure of $G(u)$ inside the unit circle is described by a finite number of simple poles,
\begin{equation}
G(u) \simeq \sum_{\alpha=1}^{r}\frac{B_\alpha}{u-z_\alpha} + G_{\rm reg}(u), \quad |z_\alpha|<1 .
\label{pole_representation}
\end{equation}
Here $G_{\rm reg}(u)$ is analytic inside the contour.
By the residue theorem, the contour moments have the form
\begin{equation}
q_n \simeq \sum_{\alpha=1}^{r}A_\alpha z_\alpha^n . 
\label{residue_prony}
\end{equation}
The amplitudes $A_\alpha$ include the residues and possible factors from the mapping. Thus the Prony form is not an extra empirical assumption on the raw moments. It follows from the pole representation of the mapped analytic function.

Equation~\eqref{residue_prony} is the key reduction. The unit-circle map turns the finite moment problem into the recovery of poles and residues of $G(u)$. The pole locations $z_\alpha$ are the Prony nodes. The residues determine the weights $A_\alpha$. If the true analytic structure contains branch cuts or other singularities, the finite pole form should be understood as an effective approximation to the part that can be resolved from the known moments.

We now continue the mapped moments by a Takagi--Prony procedure \cite{Bey0517}. The basic assumption is that the tail of $q_n$ can be locally represented by a finite exponential sum, $q_n\simeq\sum_{\alpha=1}^{M}A_\alpha z_\alpha^n$. This is equivalent to the existence of an annihilating polynomial $f(z)=\sum_{j=0}^{M}u_jz^j$, whose roots give the Prony nodes $z_\alpha$.

In the calculation, we use only the tail of the known sequence. We set $h_j=q_{n_0+j}$ with $n_0=N-2M$, and build the Hankel matrix $\mathbf H_{ab}=h_{a+b}$ for $a,b=0,\ldots,M$. For an exact $M$-mode exponential tail, this matrix has a null vector corresponding to the annihilating polynomial. With finite data, we obtain this vector from the Autonne--Takagi decomposition of $\bf H$. The Takagi vector with the smallest singular value gives the coefficients $u_j$, and the roots of $f(z)$ give the nodes $z_\alpha$.

The amplitudes $A_\alpha$ are then fitted by the least-squares scheme. The unknown mapped moments are continued as $\hat q_n=\sum_\alpha A_\alpha z_\alpha^{n-n_0}$ for $n>N$. We keep only stable roots inside the unit disk, because modes with $|z_\alpha|>1$ would generate a growing mapped tail. Finally, the continued sequence is transformed back to raw moments through the inverse of transformation \Eq{q_mu_transform}.

\begin{figure}[t]
	\centering
	\includegraphics[width=0.94\columnwidth]{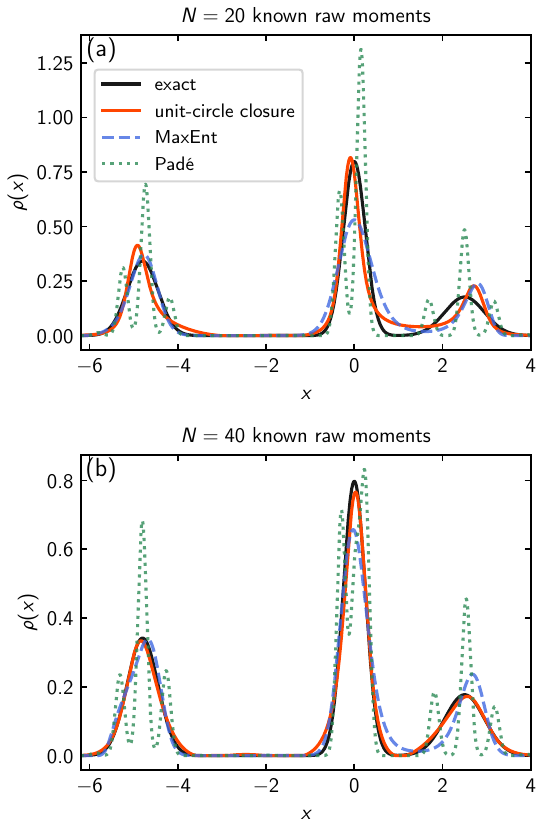}
	\caption{Comparison of density reconstruction methods from finitely many raw moments. The target density is a three-peak Gaussian mixture. The black curve is the exact density. The unit-circle closure first transforms the raw moments to unit-circle moments, continues the mapped sequence by the Takagi--Prony procedure, and then reconstructs the density. The maximum-entropy and Pad\'e moment-closure schemes are applied directly to the raw moments. Panels (a) and (b) show the results obtained from $N=20$ and $N=40$ known raw moments, respectively. The raw maximum-entropy reconstruction is stable but smooths the sharp peaks, while the Pad\'e reconstruction is more sensitive to the ill-conditioning of high-order raw moments. The unit-circle closure gives a more faithful reconstruction of the multi-peak structure.
}\label{fig2}
\end{figure}

\paragraph*{Static density reconstruction.}
We illustrate the unit-circle moment closure method with a three-peak probability density, expressed as
\[
	\rho(x) = \sum_{i=1}^3\alpha_i (2\pi\sigma_i^2)^{-1/2} \exp[{-(x-x_i)^2 / (2\sigma_i^2)}],
\]
with $\alpha_1=0.3$, $\sigma_1=0.35$, $x_1=-4.8$, $\alpha_2=0.2$, $\sigma_2=0.45$, $x_2=2.5$, $\alpha_3=0.5$, $\sigma_3=0.25$, $x_3=0$. The input data are the first $N+1$ raw moments, $\mu_0,\ldots,\mu_N$.  We compare three reconstructions in Fig.~\ref{fig2}.  The first is the present unit-circle closure.  In this case, the raw moments are first transformed to the unit-circle moments $q_n=\sum_{j=0}^{n}C_{nj}\mu_j$.  The unknown tail of $q_n$ is then continued by the Takagi--Prony procedure, and the density is reconstructed from the continued Chebyshev series.  The other two methods are standard raw-moment reconstructions.

The maximum-entropy reconstruction uses the original moment constraints directly \cite{Mea842404}. Numerically, we first rescale the variable as $y=(x-x_c)/L$ and convert the raw moments $\mu_n$ to the scaled moments $\nu_n=\langle y^n\rangle$. This rescaling is only for conditioning; the constraints are still equivalent to the raw-moment constraints. Among all positive densities satisfying these constraints, the maximum-entropy reconstruction selects the one with maximal entropy. In the scaled variable, the solution has the exponential form $\tilde\rho_{\rm ME}(y)=Z^{-1}\exp(\sum_{j=1}^{N}\lambda_j y^j)$, where the multipliers $\lambda_j$ are obtained by solving the dual convex problem. The density in the original variable is then recovered as $\rho_{\rm ME}(x)=\tilde\rho_{\rm ME}((x-x_c)/L)/L$.  This construction is stable and preserves positivity.  Its drawback is that entropy maximization favors the least structured distribution compatible with the moments, so sharp or well separated peaks are often broadened.

The Pad\'e reconstruction provides a different baseline.  It treats the raw moments as defining a rational approximation to the corresponding moment or Stieltjes transform \cite{Mea842404}.  Equivalently, it replaces the measure by a finite quadrature form, $\rho_{\text{Pad\'e}}(x)=\sum_\ell w_\ell\delta(x-x_\ell)$, whose nodes and weights are determined from the Hankel structure of the raw moments.  This approach can represent localized features more directly than maximum entropy, but it is highly sensitive to the conditioning of the raw Hankel matrices.  
% In the figure we broaden the resulting discrete measure only for visualization.

Figure\,\ref{fig2}\, (a) exhibits the result for $N=20$. The maximum-entropy curve is smooth but misses the narrow peak structure.  The Pad\'e curve is affected by the instability of the raw moments.  The unit-circle closure gives a more faithful reconstruction of the three peaks. \Fig{fig2}\,(b) shows the same comparison for $N=40$.  Increasing the number of moments improves all methods, but the same trend remains.  The result indicates that the main advantage is not simply the use of Prony continuation.  It is the Szeg\H{o} preprocessing, which moves the continuation problem from the rapidly growing raw moments to the bounded unit-circle moments.

\begin{figure}[t]
	\centering
	\includegraphics[width=0.94\columnwidth]{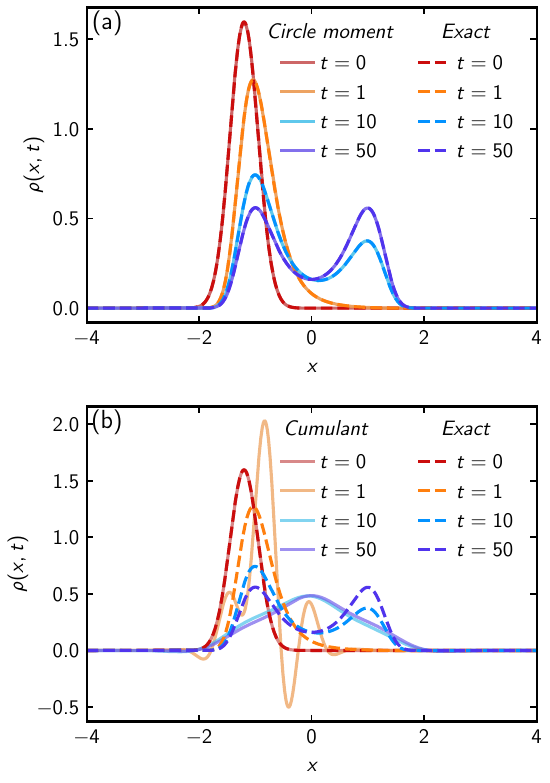}
	\caption{Time evolution of the probability density for the overdamped diffusion process in the double-well potential
\(U(x)=x^{4}/4-a x^{2}/2\), with \(a=1\) and \(D=0.2\), starting from the Gaussian state
\(\rho(x,0)\propto\exp[-(x+1.2)^2/(2\times0.25^2)]\).
Solid curves are reconstructed from the truncated moment dynamics at \(t=0,1,10,\) and \(50\), whereas dashed curves of the corresponding colors are obtained from a direct numerical solution of the Fokker--Planck equation. In the panel (a), the unit-circle closure is applied to continue the moment hierarchy, while in the panel (b), the cumulant closure is used for comparison. 
}\label{fig3}
\end{figure}

\paragraph*{Moment closure for equation of motion.} 
We consider the dynamics of an over-damped Brownian particle in a one-dimensional double-well potential, described by the Langevin equation \cite{Zwa10}
\begin{align}
	\dot x = -V'(x) + \sqrt{2D}\eta(t).
\end{align}
Here, the potential $V(x) = \frac{x^4}{4} - \frac{ax^2}{2}$ has two minima at $x=\pm\sqrt{a}$, and $\eta(t)$ is a Gaussian white noise with $\la\eta(t)\ra=0$ and $\la\eta(t)\eta(t')\ra=\delta(t-t')$. The corresponding Fokker--Planck equation for the probability density $\rho(x,t)$ is \cite{Zwa10}
\begin{align}
	\frac{\partial}{\partial t}\rho(x,t) = D\frac{\partial^2}{\partial x^2}\rho(x,t) - \frac{\partial}{\partial x}[(ax - x^3)\rho(x,t)].
\end{align}
The corresponding moment hierarchy dynamics is given by
\begin{align}
	\dot\mu_n = Dn(n-1)\mu_{n-2} + an\mu_n - n\mu_{n+2}.
\end{align}
In the numerical realization, we truncate the moment hierarchy at layer $N$. Then the $N$-th moment depends on the $(N+2)$-th moment, which is unknown. We use the unit-circle closure to continue the moment sequence and close the equations of motion. For comparison, we also apply the cumulant closure, which assumes that the cumulants of order higher than $N$ vanish. 

Figure\,\ref{fig3} compares the closed moment dynamics with the direct numerical solution of the Fokker--Planck equation. The initial Gaussian packet, localized near the left minimum, first relaxes within the left well and subsequently transfers probability across the barrier, eventually approaching the symmetric bimodal stationary distribution. For the unit-circle closure, we retain moments up to \(N_{\rm circle}=30\). As shown in \Fig{fig3} (a), the reconstructed densities remain in close agreement with the reference Fokker--Planck solution from the initial localized state to the long-time bimodal regime. The circle moment closure correctly captures both the gradual population transfer to the right well and the locations and relative weights of the two stationary peaks. No dynamical instability is observed over the time window shown.

For comparison, \Fig{fig3} (b) shows the cumulant closure method with \(N_{\rm cumulant}=10\). Although this low-order truncation can be propagated to long times, it becomes inaccurate once the distribution develops appreciable non-Gaussian and bimodal structure. Already at \(t=1\), the cumulant reconstruction exhibits spurious oscillations and a negative region, and the subsequent profiles differ substantially from the reference dynamics. Increasing the cumulant truncation order does not systematically improve the calculation: in our simulations, higher-order cumulant closures become dynamically unstable and eventually diverge. Thus, for this strongly non-Gaussian relaxation process, the unit-circle closure simultaneously provides a more accurate long-time reconstruction and a substantially more stable moment evolution than the conventional cumulant truncation.

\paragraph*{Summary.}
In this work, we introduce the unit-circle moment closure scheme, a framework for extending truncated moment hierarchies through analytic continuation rather than direct truncation. The method first transforms raw moments into unit-circle moments, for which the rapidly growing monomial basis is replaced by bounded spectral modes on the unit circle. The resulting sequence can be interpreted as contour moments of a mapped analytic function. Its resolvable singular structure is then represented by a finite set of effective poles, whose locations and amplitudes are extracted through a Takagi--Prony procedure. This construction yields a stable recurrence relation for the unresolved moment tail and permits transformation back to the original raw moments.

Several directions deserve further exploration. The present construction can be extended to multivariate distributions by mapping higher-dimensional physical domains to compact complex domains and developing corresponding multivariate spectral continuations. The unit circle can also be replaced by more general closed contours, which may provide a more flexible representation for distributions with complicated support or analytic structure. It will also be interesting to apply this idea to quantum generating functions \cite{Gu851310}, Green's functions \cite{Sil974822,Gan239187,Zha24035154,Zha25214111}, and correlation functions \cite{Liu25148001,Liu26174118}, where a finite number of moments may encode useful spectral and dynamical information. In practical settings, these moments can be obtained from quantum-chemical calculations \cite{Gan239187}, molecular dynamics simulations \cite{Bi25224106a}, or other microscopic approaches, for example for constructing realistic spectral densities. We hope that the present work offers a broader perspective on moment closure and stimulates its application to more realistic problems.

% For a multimodal Gaussian mixture, the unit-circle closure reconstructs sharp and well-separated density peaks more faithfully than the raw-moment maximum-entropy and Pad\'e reconstructions. Applied to the moment dynamics of overdamped diffusion in a double-well potential, it remains in close agreement with the direct Fokker--Planck solution from the initially localized state to the long-time bimodal regime. In contrast, the cumulant closure becomes inaccurate once strong non-Gaussianity develops, and increasing its truncation order can lead to dynamical instability. These results show that separating the conditioning problem from the continuation problem provides a practical route to stable moment closure. The present construction is naturally compatible with extensions to multivariate moment sequences, kinetic equations, stochastic reaction networks, and correlation-function hierarchies.

\paragraph*{Acknowledgments.}
Support from the National Natural Science Foundation of China (Grant Nos.\ 224B2305, 22373091) is gratefully acknowledged. The authors are indebted to Lu Han, Zi-Fan Zhu, and Wenjie Dou for invaluable discussions.

\bibliography{refs.bib}

@article{Ale13174101,
  title = {A general moment expansion method for stochastic kinetic models},
  author = {Ale, Angelique and Kirk, Paul and Stumpf, Michael P. H.},
  year = 2013,
  month = may,
  journal = {The Journal of Chemical Physics},
  volume = {138},
  number = {17},
  pages = {174101},
  issn = {0021-9606, 1089-7690},
  doi = {10.1063/1.4802475},
  urldate = {2026-06-21},
  langid = {english}
}

@book{Akh20,
  title = {The Classical Moment Problem and Some Related Questions in Analysis},
  author = {Akhiezer, N. I.},
  year = 2020,
  month = jan,
  series = {Classics in Applied Mathematics},
  publisher = {{Society for Industrial and Applied Mathematics}},
  doi = {10.1137/1.9781611976397},
  urldate = {2026-06-24},
  isbn = {978-1-61197-638-0}
}

@book{Hua87,
  title = {Statistical Mechanics},
  author = {Huang, Kerson},
  year = 1987,
  publisher = {Wiley},
  address = {New York, NY},
  isbn = {978-0-471-81518-1},
  langid = {english}
}

@incollection{Bre0287,
  title = {Pad\'e Approximations},
  booktitle = {Computational Aspects of Linear Control},
  author = {Brezinski, Claude},
  editor = {Brezinski, Claude},
  year = 2002,
  pages = {87--134},
  publisher = {Springer US},
  address = {Boston, MA},
  doi = {10.1007/978-1-4613-0261-2_4},
  urldate = {2026-06-24},
  isbn = {978-1-4613-0261-2},
  langid = {english}
}

@article{Gor68655,
  title = {Error Bounds in Equilibrium Statistical Mechanics},
  author = {Gordon, Roy G.},
  year = 1968,
  month = may,
  journal = {Journal of Mathematical Physics},
  volume = {9},
  number = {5},
  pages = {655--663},
  issn = {0022-2488},
  doi = {10.1063/1.1664624},
  urldate = {2026-06-24}
}

@article{Sta71663,
  title = {Improved Error Bounds for the Long-Range Forces between Atoms},
  author = {Starkschall, George and Gordon, Roy G.},
  year = 1971,
  month = jan,
  journal = {The Journal of Chemical Physics},
  volume = {54},
  number = {2},
  pages = {663--673},
  issn = {0021-9606},
  doi = {10.1063/1.1674894},
  urldate = {2026-06-24}
}

@article{Gri12154105,
  title = {A study of the accuracy of moment-closure approximations for stochastic chemical kinetics},
  author = {Grima, Ramon},
  year = 2012,
  month = apr,
  journal = {The Journal of Chemical Physics},
  volume = {136},
  number = {15},
  pages = {154105},
  issn = {0021-9606},
  doi = {10.1063/1.3702848},
  urldate = {2026-06-24}
}

@article{Liu26174118,
  title = {Stable memory kernel coupling theory for quantum dynamics: Projection-based and continued fraction methods},
  shorttitle = {Stable memory kernel coupling theory for quantum dynamics},
  author = {Liu, Wei and Bi, Rui-Hao and Su, Yu and Xu, Limin and Zhou, Zhennan and Wang, Yao and Dou, Wenjie},
  year = 2026,
  month = may,
  journal = {The Journal of Chemical Physics},
  volume = {164},
  number = {17},
  pages = {174118},
  issn = {0021-9606},
  doi = {10.1063/5.0327266},
  urldate = {2026-06-02}
}

@article{Bey0517,
  title = {On approximation of functions by exponential sums},
  author = {Beylkin, Gregory and Monz{\'o}n, Lucas},
  year = 2005,
  month = jul,
  journal = {Applied and Computational Harmonic Analysis},
  volume = {19},
  number = {1},
  pages = {17--48},
  issn = {1063-5203},
  doi = {10.1016/j.acha.2005.01.003},
  urldate = {2026-06-27}
}

@book{Sim05,
  title = {Orthogonal Polynomials on the Unit Circle: Part 1: Classical Theory},
  shorttitle = {Orthogonal Polynomials on the Unit Circle},
  author = {Simon, Barry},
  year = 2005,
  publisher = {American Mathematial Society},
  address = {Providence, Rhode Island},
  isbn = {978-0-8218-4863-0},
  langid = {english}
}

@book{Sim13,
  title = {Orthogonal Polynomials on the Unit Circle: Part 2: Spectral Theory},
  shorttitle = {Orthogonal Polynomials on the Unit Circle},
  author = {Simon, Barry},
  year = 2013,
  publisher = {American Mathematial Society},
  address = {Providence},
  isbn = {978-0-8218-4864-7}
}

@book{Zwa10,
  title = {Nonequilibrium Statistical Mechanics},
  author = {Zwanzig, Robert},
  year = 2010,
  address = {Oxford  u.a.},
  isbn = {978-0-19-514018-7}
}

@article{Gan239187,
  title = {Doubles Connected Moments Expansion: A Tractable Approximate Horn--Weinstein Approach for Quantum Chemistry},
  shorttitle = {Doubles Connected Moments Expansion},
  author = {Ganoe, Brad and {Head-Gordon}, Martin},
  year = 2023,
  month = dec,
  journal = {Journal of Chemical Theory and Computation},
  volume = {19},
  number = {24},
  pages = {9187--9201},
  issn = {1549-9618, 1549-9626},
  doi = {10.1021/acs.jctc.3c00929},
  urldate = {2026-06-21},
  copyright = {https://creativecommons.org/licenses/by/4.0/},
  langid = {english}
}

@article{Bi25224106a,
  title = {Universal structure of computing moments for exact quantum dynamics: Application to arbitrary system--bath couplings},
  shorttitle = {Universal structure of computing moments for exact quantum dynamics},
  author = {Bi, Rui-Hao and Liu, Wei and Dou, Wenjie},
  year = 2025,
  month = jun,
  journal = {The Journal of Chemical Physics},
  volume = {162},
  number = {22},
  pages = {224106},
  issn = {0021-9606},
  doi = {10.1063/5.0273707},
  urldate = {2026-06-27}
}

@article{Gu851310,
  title = {Group-theoretical formalism of quantum mechanics based on quantum generalization of characteristic functions},
  author = {Gu, Yan},
  year = 1985,
  journal = {Physical Review A},
  volume = {32},
  number = {3},
  pages = {1310},
  publisher = {APS}
}

@article{Zha24035154,
  title = {Minimal pole representation and controlled analytic continuation of Matsubara response functions},
  author = {Zhang, Lei and Gull, Emanuel},
  year = 2024,
  month = jul,
  journal = {Physical Review B},
  volume = {110},
  number = {3},
  pages = {035154},
  publisher = {American Physical Society},
  doi = {10.1103/PhysRevB.110.035154},
  urldate = {2026-06-24}
}

@article{Liu25148001,
  title = {Memory Kernel Coupling Theory: Obtaining Time Correlation Function from Higher-Order Moments},
  shorttitle = {Memory Kernel Coupling Theory},
  author = {Liu, Wei and Su, Yu and Wang, Yao and Dou, Wenjie},
  year = 2025,
  month = sep,
  journal = {Physical Review Letters},
  volume = {135},
  number = {14},
  pages = {148001},
  issn = {0031-9007, 1079-7114},
  doi = {10.1103/qvd5-5z6m},
  urldate = {2026-06-01},
  langid = {english}
}

@article{Alv26025207,
  title = {High-order moment closure for nonmagnetized electrons in partially ionized plasmas},
  author = {Alvarez Laguna, A. and Hara, K.},
  year = 2026,
  month = feb,
  journal = {Physical Review E},
  volume = {113},
  number = {2},
  pages = {025207},
  issn = {2470-0045, 2470-0053},
  doi = {10.1103/lm1z-bzt3},
  urldate = {2026-06-23},
  langid = {english}
}

@article{Fro25063022,
  title = {Quantum maximum entropy closure for small flavor coherence},
  author = {Froustey, Julien and Kneller, James P. and McLaughlin, Gail C.},
  year = 2025,
  month = mar,
  journal = {Physical Review D},
  volume = {111},
  number = {6},
  pages = {063022},
  issn = {2470-0010, 2470-0029},
  doi = {10.1103/PhysRevD.111.063022},
  urldate = {2026-06-24},
  langid = {english}
}

@article{Gra49331,
  title = {On the kinetic theory of rarefied gases},
  author = {Grad, Harold},
  year = 1949,
  month = dec,
  journal = {Communications on Pure and Applied Mathematics},
  volume = {2},
  number = {4},
  pages = {331--407},
  issn = {0010-3640, 1097-0312},
  doi = {10.1002/cpa.3160020403},
  urldate = {2026-06-24},
  langid = {english}
}

@article{Ott10120601,
  title = {Thermodynamically Admissible 13 Moment Equations from the Boltzmann Equation},
  author = {{\"O}ttinger, Hans Christian},
  year = 2010,
  month = mar,
  journal = {Physical Review Letters},
  volume = {104},
  number = {12},
  pages = {120601},
  issn = {0031-9007, 1079-7114},
  doi = {10.1103/PhysRevLett.104.120601},
  urldate = {2026-06-23},
  copyright = {http://link.aps.org/licenses/aps-default-license},
  langid = {english}
}

@article{Ran06178101,
  title = {Maximum-Entropy Closures for Kinetic Theories of Neuronal Network Dynamics},
  author = {Rangan, Aaditya V. and Cai, David},
  year = 2006,
  month = may,
  journal = {Physical Review Letters},
  volume = {96},
  number = {17},
  pages = {178101},
  issn = {0031-9007, 1079-7114},
  doi = {10.1103/PhysRevLett.96.178101},
  urldate = {2026-06-23},
  copyright = {http://link.aps.org/licenses/aps-default-license},
  langid = {english}
}

@article{Bli98193,
  title = {Expansions for nearly Gaussian distributions},
  author = {Blinnikov, S. and Moessner, R.},
  year = 1998,
  month = may,
  journal = {Astronomy and Astrophysics Supplement Series},
  volume = {130},
  number = {1},
  pages = {193--205},
  issn = {0365-0138, 1286-4846},
  doi = {10.1051/aas:1998221},
  urldate = {2026-06-07},
  langid = {english}
}

@article{Flo20557277,
  title = {Recent Advances in the Calculation of Dynamical Correlation Functions},
  author = {Florencio, J. and De Alcantara Bonfim, O. F.},
  year = 2020,
  month = nov,
  journal = {Frontiers in Physics},
  volume = {8},
  pages = {557277},
  issn = {2296-424X},
  doi = {10.3389/fphy.2020.557277},
  urldate = {2026-06-21},
  langid = {english}
}

@article{Sil974822,
  title = {Calculation of densities of states and spectral functions by Chebyshev recursion and maximum entropy},
  author = {Silver, R. N. and R{\"o}der, H.},
  year = 1997,
  month = oct,
  journal = {Physical Review E},
  volume = {56},
  number = {4},
  pages = {4822--4829},
  issn = {1063-651X, 1095-3787},
  doi = {10.1103/PhysRevE.56.4822},
  urldate = {2026-06-27},
  copyright = {http://link.aps.org/licenses/aps-default-license},
  langid = {english}
}

@book{Kue15,
  title = {Multiple Time Scale Dynamics},
  author = {Kuehn, Christian},
  year = 2015,
  series = {Applied Mathematical Sciences},
  volume = {191},
  publisher = {Springer International Publishing},
  address = {Cham},
  doi = {10.1007/978-3-319-12316-5},
  urldate = {2026-06-21},
  copyright = {https://www.springernature.com/gp/researchers/text-and-data-mining},
  isbn = {978-3-319-12315-8 978-3-319-12316-5},
  langid = {english}
}

@book{Kue16,
  title = {Moment Closure - A Brief Review},
  author = {Kuehn, Christian},
  year = 2016,
  eprint = {1505.02190},
  primaryclass = {cond-mat.stat-mech},
  doi = {10.1007/978-3-319-28028-8},
  urldate = {2026-06-21},
  archiveprefix = {arXiv}
}

@article{Lev961021,
  title = {Moment closure hierarchies for kinetic theories},
  author = {Levermore, C. David},
  year = 1996,
  month = jun,
  journal = {Journal of Statistical Physics},
  volume = {83},
  number = {5-6},
  pages = {1021--1065},
  issn = {0022-4715, 1572-9613},
  doi = {10.1007/BF02179552},
  urldate = {2026-06-21},
  copyright = {http://www.springer.com/tdm},
  langid = {english}
}

@article{Mea842404,
  title = {Maximum entropy in the problem of moments},
  author = {Mead, Lawrence R. and Papanicolaou, N.},
  year = 1984,
  month = aug,
  journal = {Journal of Mathematical Physics},
  volume = {25},
  number = {8},
  pages = {2404--2417},
  issn = {0022-2488, 1089-7658},
  doi = {10.1063/1.526446},
  urldate = {2026-06-07},
  langid = {english}
}

@article{Sch15185101,
  title = {Comparison of different moment-closure approximations for stochastic chemical kinetics},
  author = {Schnoerr, David and Sanguinetti, Guido and Grima, Ramon},
  year = 2015,
  month = nov,
  journal = {The Journal of Chemical Physics},
  volume = {143},
  number = {18},
  pages = {185101},
  issn = {0021-9606, 1089-7690},
  doi = {10.1063/1.4934990},
  urldate = {2026-06-21},
  langid = {english}
}

@article{Wuy22054312,
  title = {Mean-field models of dynamics on networks via moment closure: An automated procedure},
  shorttitle = {Mean-field models of dynamics on networks via moment closure},
  author = {Wuyts, Bert and Sieber, Jan},
  year = 2022,
  month = nov,
  journal = {Physical Review E},
  volume = {106},
  number = {5},
  pages = {054312},
  issn = {2470-0045, 2470-0053},
  doi = {10.1103/PhysRevE.106.054312},
  urldate = {2026-06-21},
  langid = {english}
}

@article{Zha25214111,
  title = {Minimal pole representation for spectral functions},
  author = {Zhang, Lei and Erpenbeck, Andr{\'e} and Yu, Yang and Gull, Emanuel},
  year = 2025,
  month = jun,
  journal = {The Journal of Chemical Physics},
  volume = {162},
  number = {21},
  pages = {214111},
  issn = {0021-9606},
  doi = {10.1063/5.0273763},
  urldate = {2025-09-24}
}

\end{document}